\documentclass[%
 reprint,
superscriptaddress,
showpacs,preprintnumbers,
nofootinbib,
 amsmath,amssymb,aps,
prd,
floatfix,
]{revtex4-1}

\usepackage{lipsum} 

\usepackage{bm} 
\usepackage{physics}
\usepackage{comment}
\pdfoutput=1
\usepackage{amsmath,amsfonts,amsthm}
\usepackage{esdiff} 
\usepackage{booktabs}  
\usepackage{url}  
\usepackage[hidelinks]{hyperref}  
\usepackage{relsize}

\usepackage{cleveref}  
	\crefname{equation}{equation}{equations}
	\crefname{figure}{figure}{figures}	
	\crefname{table}{table}{tables}
\usepackage[caption=false]{subfig}

\usepackage[normalem]{ulem}

\usepackage[usenames,dvipsnames,svgnames,table]{xcolor}

\usepackage[export]{adjustbox}
\usepackage{graphicx}

\usepackage[sc]{mathpazo} 
\usepackage[T1]{fontenc} 
\usepackage{microtype} 

\usepackage{booktabs} 
\usepackage{float} 
\usepackage{hyperref} 

\usepackage{lettrine} 
\usepackage{paralist} 

\usepackage{titlesec} 
\renewcommand\thesection{\Roman{section}} 
\renewcommand\thesubsection{\Alph{subsection}} 
\titleformat{\section}[block]{\large\scshape\centering\bfseries}{\thesection.}{1em}{} 

\titleformat{\subsection}[block]{\scshape\centering}{\thesubsection.}{1em}{} 

\DeclareCaptionFormat{myformat}{#1#2#3\hrulefill}
\begin{document}

\title{Inference finds consistency between a neutrino flavor evolution model and Earth-based solar neutrino measurements} 

\author{Caroline Laber-Smith}\thanks{Corresponding author}
\email{labersmith@wisc.edu}
\affiliation{Department of Physics, University of Wisconsin: Madison, Madison, WI 53706, USA}
\author{A.A. Ahmetaj}
\email{aahmet03@nyit.edu}
\affiliation{Department of Physics, New York Institute of Technology, New York, NY 10023, USA}
\author{Eve Armstrong}
\email{earmst01@nyit.edu}
\affiliation{Department of Physics, New York Institute of Technology, New York, NY 10023, USA}
\affiliation{Department of Astrophysics, American Museum of Natural History, New York, NY 10024, USA}
\author{A. Baha Balantekin}
\email{baha@physics.wisc.edu}
\affiliation{Department of Physics, University of Wisconsin: Madison, Madison, WI 53706, USA}
\author{Amol V.\ Patwardhan}
\email{apatward@slac.stanford.edu} 
\affiliation{SLAC National Accelerator Laboratory, Menlo Park, CA 94025, USA}
\author{M. Margarette Sanchez}
\email{msanch11@nyit.edu}
\affiliation{Department of Physics, New York Institute of Technology, New York, NY 10023, USA}
\author{Sherry Wong}
\email{bwong24@wisc.edu}
\affiliation{Department of Astronomy, University of Wisconsin: Madison, Madison, WI 53706, USA}
\date{\today}

\begin{abstract}
We continue examining statistical data assimilation (SDA), an inference methodology, to infer solutions to neutrino flavor evolution, for the first time using real -- rather than simulated -- data.  The model represents neutrinos streaming from the Sun's center and undergoing a Mikheyev-Smirnov-Wolfenstein (MSW) resonance in flavor space, due to the radially-varying electron number density.  The model neutrino energies are chosen to correspond to experimental bins in the Sudbury Neutrino Observatory (SNO) and Borexino experiments, which measure electron-flavor survival probability at Earth.  The procedure successfully finds consistency between the observed fluxes and the model, if the MSW resonance -- that is, flavor evolution due to solar electrons -- is included in the dynamical equations representing the model.
\end{abstract}

\preprint{N3AS-22-022, SLAC-PUB-17705}

\maketitle 
\maketitle 


\section{Introduction}

Neutrinos are ubiquitous in astrophysical environments.  In stars like the Sun, they are a byproduct of the nuclear reactions that provide the energy flux counteracting gravity.  Proto-neutron stars formed following the core collapse in a supernova cool by emitting copious neutrino pairs, and similar neutrino emission takes place following neutron star mergers.  Observations of those neutrinos would provide not only valuable information about the interior of the star and collapse mechanism, but also the nature and properties of neutrinos.  In particular, neutrino "flavor," a property that dictates neutrino interaction with matter, significantly affects the physics of these events~\cite{Balantekin:1988aq,Fuller:1992eu,Qian:1993dg,Fuller:1993ry,Fuller:1995qy,Balantekin:1998yb,Duan:2010af,Wu:2014kaa,Wu:2015glr,Sasaki:2017jry,Balantekin:2017bau,Balantekin:2018mpq,Xiong:2019nvw,Xiong:2020ntn}.

A complete description of neutrino flavor evolution and transport in astrophysical environments is technically very involved.  Powerful numerical integration codes exist for obtaining solutions to the flavor evolution problem in compact object environments~\cite{duan2006simulation,duan2008simulating,richers2019neutrino,Richers:2021nbx,Richers:2021xtf,Kato:2021cjf,Nagakura:2022qko,George:2022lwg}. Many of these codes, however, require adopting rather rigid physical assumptions regarding the symmetries of the problem, and it has been shown in recent years that relaxing these assumptions reveals physics that had been artificially hidden (e.g., see Ref.~\cite{tamborra2021new,Richers:2022zug} and references therein).  Further, because of the large density of neutrinos in these regimes, neutrino-neutrino interactions introduce nonlinear elements in the transport, rendering the problem computationally taxing for the existing large-scale codes. Additionally, these nonlinear "collective oscillations" could exhibit signatures of many-body quantum correlations among neutrinos (e.g.,~\cite{Patwardhan:2019zta,Cervia:2019res,Rrapaj:2019pxz,Patwardhan:2021rej,Roggero:2021asb,Xiong:2021evk,Martin:2021bri,Roggero:2022hpy,Cervia:2022pro}). Hence it is helpful to explore the suitability of different computational tools to treat neutrino flavor evolution and transport.

In previous papers{ ~\cite{armstrong2022inferenceA,armstrong2022inferenceB,armstrong2017optimization,armstrong2020inference,rrapaj2021inference}, we applied} an inference technique -- a fundamentally different framework compared to forward integration -- to examine nonlinear collective oscillations in core-collapse supernovae (CCSN). Inference is a means to optimize a model given measurements, where measurements are assumed to arise from model dynamics. 
 
The specific technique used is statistical data assimilation (SDA), which was invented for numerical weather prediction~\cite{kimura2002numerical,kalnay2003atmospheric,evensen2009data,betts2010practical,whartenby2013number,an2017estimating} for the case of sparse data.  It has since gained considerable traction in neurobiology~\cite{schiff2009kalman, toth2011dynamical,kostuk2012dynamical,hamilton2013real,meliza2014estimating,nogaret2016automatic,armstrong2020statistical}, and within astrophysics has been applied to exoplanet modeling~\cite{Madhusudhan2018} and solar cycle prediction~\cite{Kitiashvili2008,kitiashvili2020}.

%
 
{ To date, we have applied SDA to small-scale neutrino flavor evolution models, using simulated data.  Our first work~\cite{armstrong2017optimization} established that SDA is capable -- in principle -- of finding solutions to the flavor evolution problem, given Earth-based measurements together with assumptions regarding flavor states at emission.  Next, in Ref.~\cite{armstrong2020inference} we devised a litmus test for identifying correct solutions, and Ref.~\cite{rrapaj2021inference} compared the efficacy of our SDA technique to alternative (e.g. "neural differential equations") approaches.  More recently, we challenged SDA to solve problems that forward integration renders difficult to access: 1) direction-changing scattering in the model dynamics~\cite{armstrong2022inferenceB}, and 2) unknown initial conditions of the neutrino flavor field at emission from the CCSN core~\cite{armstrong2022inferenceA}.}



 In this paper we seek, for the first time, to test the SDA procedure using real neutrino data from experiments.  Data on neutrinos from CCSN are limited to one event: SN1987A~\cite{Kamiokande:1987a,IMB:1987a,Baksan:1987a}, wherein about 20 neutrinos were detected -- not a sufficiently large sample to robustly investigate flavor evolution signatures.  Hence we turn to an environment from which neutrino data are copious: the Sun.  Our purpose here is to explore the efficacy of inference to describe the matter-enhanced neutrino oscillations in the Sun.  Furthermore, it is worthwhile to investigate possible improvements that inference might offer for the analysis of solar neutrino data.  In particular, inference might offer an independent check on new methods to remove cosmogenic-induced spallation in Super-Kamiokande, a recent effort to improve the precision of solar neutrino data~\cite{Super-Kamiokande:2021snn}.

To that end, we challenge the SDA procedure to find a solution -- in terms of neutrino flavor evolution with radius, in a steady-state model -- that is consistent with three elements: the standard solar model, a model of neutrino flavor evolution, and observed neutrino fluxes at Earth.  In addition, we explore the significantly more complex problem of state prediction simultaneous with the estimation of unknown model parameters.

\section{Input} 

\subsection{Neutrino flavor evolution} \label{sec:evol}

We consider two-flavor mixing of solar neutrinos, which is a good approximation because of the small value of the mixing angle $\theta_{13}$ \cite{Balantekin:2003dc}. Hence the angle $\theta_{12}$ describes the effective mixing between the electron neutrinos, $\psi_e$, and a linear combination of the muon and tau neutrinos, $\psi_x$.  Hereafter we set $\theta = \theta_{12}$, dropping the subscripts.  We assume that flavor evolution is driven entirely by coherent forward-scattering.  This arises from neutrino-matter interactions, and leads to an in-medium effective neutrino mass level crossing, referred to as the \lq\lq MSW resonance.\rq\rq~\cite{mikheev1985resonance,Mikheev:1986if,wolfenstein1978neutrino}. The MSW resonance is associated with an enhanced $e \leftrightarrow x$ flavor conversion probability.  The neutrinos are produced at the core of the Sun as electron neutrinos, then they undergo flavor oscillations while also forward-scattering off the background electrons as they travel through the Sun. For each neutrino, this propagation is described either by the equation: 
\begin{widetext}
\begin{equation}
\label{model2}    
i \diff{}{r}  \left(\begin{array}{c}
     \psi_e \\
  {\psi_x} 
\end{array} \right)  = \frac{\omega}{2} 
\left( \begin{array}{cc}
   \frac{V(r)}{\omega} - \cos 2 \theta & \sin 2 \theta \\
 \sin 2 \theta    &  - \frac{V(r)}{\omega} + \cos 2 \theta 
\end{array} \right) 
 \left(\begin{array}{c}
     \psi_e \\
  {\psi_x}  
\end{array} \right)
\end{equation}
\end{widetext}
\noindent
or by the equation~\cite{Sigl1993,Loreti:1994ry}: 
\begin{equation} 
\label{eq:model1}
\diff{\vec{P}}{r} = \left(\omega \vec{B} + V(r) \hat{z}\right) \times \vec{P},
\end{equation}
\noindent where $\omega = \delta m^2/(2E)$ is the vacuum oscillation frequency of a neutrino with energy $E$, $\delta m^2$ being the mass-squared difference in vacuum.  The unit vector $\vec{B}=\sin(2 \theta) \hat{x} -\cos(2 \theta) \hat{z}$ represents flavor mixing in vacuum, with mixing angle $\theta$.  The function $V(r)$ arises from neutrino interactions with the background electrons.  Assuming that flavor evolution is driven entirely by coherent forward-scattering, $V(r)$ is given by the Wolfenstein correction to the neutrino mass: $ V(r) = \sqrt{2} G_F N_e(r)$~\cite{wolfenstein1978neutrino}, where $N_e$ is the electron density, taken from the Standard Solar Model~\cite{bahcall2005new}. Finally, the relation between the dynamical variables in Eq.~(\ref{model2}) and Eq.~(\ref{eq:model1}) is: 
\begin{equation}
\label{polvec}
\left(
    \begin{array}{c}
      P_x     \\
      P_y   \\
      P_z
    \end{array} \right) = \left(
    \begin{array}{c}
    \psi_e \psi_x^* + \psi_e^* \psi_x \\
    i( \psi_e \psi_x^* - \psi_e^* \psi_x) \\
         |\psi_e|^2 - |\psi_x|^2 \\
             \end{array} \right) .
\end{equation}

In Eq.~(\ref{eq:model1}), the flavor state of the neutrino is characterized by the \lq\lq polarization vector\rq\rq\ $\vec{P}$. Eq.~(\ref{polvec}) shows that the components of the polarization vector are real numbers.  In the inference method adopted in this work, the dynamical variables and input parameters need to real.  For that reason, in this paper we adopt Eq.~(\ref{eq:model1}) to describe neutrino evolution. The density matrix $\rho$ of the system can also be written in terms of the polarization vector as:
\begin{equation}
\label{dmatrix}
    \rho = \frac12 \left(\mathbb{1}+\vec\sigma \cdot \vec P \right) .
\end{equation}

Here, the $\hat{z}$ component of the neutrino polarization vector denotes the net flavor content of electron flavor minus the superposition of muon and tau flavors.  Hence in the Sun, Eq. (\ref{model2}) is solved with the initial conditions $\psi_e(0) = 1$, $\psi_x(0) = 0$, or equivalently, Eq. (\ref{eq:model1}) is solved with the initial conditions $P_z= +1, P_x = P_y =0$. In this configuration, the density matrix as defined in Eq. (\ref{dmatrix}) represents a pure quantum state. In the forward-scattering regime, this purity (i.e. $\rho^2 = \rho$) is preserved through the course of flavor evolution, and one therefore has $|\vec{P}|^2=1$ throughout the evolution. 

We solve Eq. (\ref{eq:model1}) inside the Sun, using the inference procedure to be described in later sections.  In particular, the solution includes a prediction of $\vec{P}$ at the surface of the Sun.  As our model does not extend beyond the Sun, we need to use the time evolution operator to relate \(\vec{P}\) at the endpoint of our model (i.e. the solar surface) to the measured electron flavor survival probability, \(P_{\text{survival}}\), at Earth's surface\footnote{We note one consideration that was not made in this paper, but which might be important in future work.  In calculating this evolution over long distances during which no measurements are made, it might be important to consider the phenomenon of neutrino state decoherence due to spatial separation of mass eigenstates; see Appendix for more details.}.

Neutrino propagation from the solar surface (denoted by \lq $\odot s$\rq) to Earth (denoted by \lq $\oplus$\rq) can be analytically calculated to obtain the density matrix on Earth: 
\begin{equation}
\frac12 U(L) \left(1+\vec{\sigma}\vdot\vec{P}_{\odot s}\right)U^\dagger(L) = \frac12 \left(1+\vec{\sigma}\vdot \vec{P}_\oplus\right),
\end{equation}
where the operator \(U\) is:
\begin{equation}
\label{umatrix}
U = \begin{pmatrix}\psi_e\left(L\right) && -\psi^*_x \left(L\right) \\ \psi_x \left(L\right) && \psi^*_e\left(L\right) \end{pmatrix} .
\end{equation}

In Eq. (\ref{umatrix}) the entries are calculated by solving Eq.~(\ref{model2}) in vacuum: 
\begin{gather}
\psi_e(L) = \cos \omega L - i \cos 2 \theta \sin \omega L , \\
\psi_x(L) = i \sin 2\theta \sin \omega L.
\end{gather} 
where $L$ is the Earth-Sun distance. If one uses only the \lq day\rq\ data (so that the measured neutrinos haven't passed through the earth), as we do, then $\vec{P}_\oplus$ represents the flavor state of the neutrino at the detector. 
Hence for the $z$ component of \(\vec P_\oplus\) on or near the surface of Earth, we have:
\begin{widetext}
\begin{equation}
P_{\oplus,z} = \left(1-2\sin^2 2\theta \sin^2 \omega L\right) P_{\odot s,z} \\- \left(2\cos2\theta \sin2\theta \sin^2\omega L\right) P_{\odot s,x} - \frac12 \left(\sin2\omega L \sin2\theta\right)P_{\odot s,y}. 
\end{equation}
\end{widetext}
Averaging over \(\omega L\) (that is,  over multiple oscillation cycles in vacuum), we obtain:
\begin{equation}
\label{eq:Pzearth}
P_{\oplus,z} = \left(1-\sin^2 2\theta\right) P_{\odot s, z} - \cos2\theta \sin2\theta \, P_{\odot s, x}.
\end{equation}

\(P_{\oplus,z}\) can be related to the measured \(P_{\text{survival}}\) using:
\begin{equation} \label{eq:PzPs}
P_{\oplus,z} = 2 P_{\text{survival}}-1,
\end{equation}
and hence we have the following constraint on the \lq final\rq\ (i.e., solar surface) values of our state variables:
\begin{equation}
\label{eq:connection}
\left(1-\sin^2 2\theta\right) P_{\odot s,z} - \cos2\theta \sin2\theta P_{\odot s,x} = 2 P_{\text{survival}}-1.
\end{equation}
\noindent
Eq. (\ref{eq:connection}) yields a linear relation between \(P_{z}\) and \(P_x\) at the surface of the Sun.

\subsection{Neutrino data} 
To test the inference procedure, we use $^8$B day-time neutrino flux observed by the SNO \cite{SNO:2009uok} and Borexino \cite{BOREXINO:2018ohr} experiments. 

For the Borexino data, we use only the observed pp-chain neutrinos, and not the {carbon-nitrogen-oxygen cycle (CNO)} neutrinos. This is a reasonable choice for the Sun, as its core temperature is relatively low, meaning that few CNO neutrinos are produced.  In addition, for simplicity we use day data only.  The Borexino survival probabilities are listed for three discrete energies~\cite{BOREXINO:2018ohr}.

The SNO collaboration used an analytic fitting formula for the survival probability \(P_\text{survival}\)~\cite{SNO:2009uok} as a function of neutrino energy $E_\nu$.  For daytime data only, the formula is:
\begin{equation} \label{eq:SNO}
\begin{split}
P&_\text{survival} \left(E_\nu\right) \\
&= c_0 + c_1\left(E_\nu-10\,\text{MeV}\right) + c_2 \left(E_\nu-10\, \text{MeV}\right)^2,
\end{split}
\end{equation}
\noindent where the best-fit values of the coefficients $c_0, \ldots, c_2$, along with the uncertainties, are given in Table~\ref{table:params}. For the neutrino oscillation parameters $\delta m^2$ and $\theta$, we use the following values:
\begin{gather}
    \delta m^2 = 7.530\times10^{-17} \text{ MeV}^2\\
    \theta = 0.5838 \text{ radians}
\end{gather}

\section{Inference Methodology} \label{sec:method}

\subsection{General formulation} \label{sec:methodGen}

Statistical data assimilation is an inference procedure wherein measured quantities are assumed to arise from a dynamical physical model.  It is designed for cases wherein only a subset of the model state variables can be experimentally accessed.  We write this model $\bm{F}$ in $D$ ordinary differential equations:
\begin{equation} \label{eq:ODE}
  \diff{x_a(r)}{r} = F_a(\bm{x}(r),\bm{p}(r)); \hspace{1em} a =1,2,\ldots,D,
\end{equation}
\noindent
where $r$ is the parameterization -- for example, distance or time.  Components $x_a$ of vector $\bm{x}$ are the model state variables.  The $\bm{p}$ are any unknown parameters to be estimated, and they may vary with $r$.

Measured quantities comprise a subset $L$ of the $D$ state variables.  We seek to estimate the evolution of all state variables that is consistent with measurements, and to predict model evolution at parameterized locations where measurements do not exist.  

\subsection{Optimization formulation} \label{sec:methodOpt}

We use a path-integral formulation of SDA, which can be summarized in three equations.  The path integral is an integral representation of the master equation for the stochastic process represented by Eq.~(\ref{eq:ODE}).  We seek the probability of obtaining a path $\bm{X}$ in the model's state space given observations $\bm{Y}$: 
\begin{align}
  P(\bm{X}|\bm{Y}) = e^{-A_0(\bm{X},\bm{Y})}.
\end{align}
\noindent
This expression states: \textit{the path $\bm{X}$ for which the probability - given $\bm{Y}$ - is greatest is the path that minimizes the quantity $A_0$}, which we call our action.  A formulation for $A_0$ will permit us to obtain the expectation value of any function $G(\bm{X})$ on a path $\bm{X}$: 
\begin{align}
  G(\bm{X}) = \langle G(\bm{X}) \rangle = \frac{\int d\bm{X} G(\bm{X}) e^{-A_0(\bm{X},\bm{Y})}}{\int d\bm{X} e^{-A_0(\bm{X},\bm{Y})}}.  
\end{align}
\noindent
Expectation values are the quantities of interest when the problem is statistical in nature.  For many estimation problems, the quantity of interest is the path itself: $G(\bm{X}) = \bm{X}$.  The action is written in two terms:
\begin{equation}
\begin{aligned}
\label{eq:action}
A_0(\bm{X},\bm{Y}) &= -\mathlarger{\sum} \log[P(\bm{x}(n+1)|\bm{x}(n))]\\ 
    &-\mathlarger{\sum} \text{CMI}(\bm{x}(n),\bm{y}(n)|\bm{Y}(n-1)).
\end{aligned}
\end{equation}

The first term describes Markov-chain transition probabilities governing the model dynamics.  The second term is the conditional mutual information (CMI) ~\cite{fano1961transmission}, which asks: "How
much information, in bits, is learned about event $\bm{x}$(n)  upon observing event $\bm{y}$(n), conditioned on having previously observed event(s) $\bm{Y}$(n - 1)?”\footnote{The measurement term can be considered a synchronization term, which are often introduced artificially into control problems. Here, however, the measurement term arises naturally through considering the effects of the information those measurements contain.}.  Simplifications are then made to write a computationally-functional form of $A_0$, and model-specific equality constraints may be added.  See Ref.~\cite{abarbanel2013predicting} for a derivation of Eq.~\eqref{eq:action}.

The SDA problem is then cast as an optimization, where the action is a cost function - a succinct and powerful equivalency.  The cost function of the optimizer is equivalent to the action on paths in the state space that is searched.  Generally, the action surface is $((D + p) \times (N+1))$-dimensional, where $N+1$ is the number of discretized model locations, taken to be independent dimensions.  One seeks the path $\bm{X}^0 = \{ \bm{x}(0),\ldots,\bm{x}(N),\bm{p}(0),\ldots,\bm{p}(N) \}$ in state space on which $A_0$ attains a minimum value.  Minima are found by requiring that small variations to the action vanish under small perturbations~\cite{oden2012variational}, thereby enforcing the Euler-Lagrange equations of motion upon any path.  We extremize the cost function via the variational method.  

After many simplifications (see Appendix A of Ref~\cite{armstrong2017optimization}), the Markov-chain term (first term of Eq.~(\ref{eq:action})) reduces to a \lq\lq model error\rq\rq, which describes the divergence of the prediction from model dynamics.  The mutual information (second term of Eq.~(\ref{eq:action})) reduces to a \lq\lq measurement error\rq\rq, describing the divergence of the prediction from measurements. (For a pedagogical treatment, see Ref.~\cite{abarbanel2013predicting}.)  The action $A_0$ used in this paper is written as:
\begin{widetext}
\begin{equation} \label{eq:actionlong}
\begin{split}
A_0 =& R_f A_\text{model} + R_m A_\text{meas}\\
A_\text{model}=&\frac{1}{{N}D}	\mathlarger{\sum}_{n \in \{\text{odd}\}}^{N-2} \, \mathlarger{\sum}_{a=1}^D \\ 
   & \Bigg[ \left\{x_a(r_{n+2}) - x_a(r_n) - \frac{\delta r}{6} [F_a(\bm{x}(r_n), \bm{p}(r_n)) + 4F_a(\bm{x}(r_{n+1}),\bm{p}(r_{n+1})) + F_a(\bm{x}(r_{n+2}),\bm{p}(r_{n+2}))]\right\}^2 \\
   & + \left\{ x_a(r_{n+1}) - \frac12 \left(x_a(r_n)+x_a(r_{n+2})\right) - \frac{\delta r}{8} [F_a(\bm{x}(r_n),\bm{p}(r_n)) - F_a(\bm{x}(r_{n+2}),\bm{p}(r_{n+2)})]\right\}^2 \Bigg] \\
  A_{\text{meas}} =& \frac{1}{N_{\text{meas}}} \mathlarger{\sum}_{r_m \in \{\text{meas}\}} \, \mathlarger{\sum}_{l=1}^d  \left[\left(y_{l}\left(r_m\right) - h_{l,m}(\bm{x}(r_m) \right)^2 \right]
\end{split}
\end{equation}
\end{widetext}

The model error, $A_\text{model}$, imposes adherence to the model evolution of all $D$ state variables $x_a$.  The outer sum on $n$ runs through all odd-numbered discretized locations.  The sum on $a$ runs through all $D$ state variables. The terms within the first and second sets of curly brackets represent the errors in the first and second derivatives, respectively, of the state variables.



The measurement error, $A_\text{meas}$, imposes adherence to measurements. The variables $y_l$, for $l=1,\ldots,d$, represent the $d$ quantities measured at locations $r_m \in \{{\text{meas}}\}$, where $N_\text{meas}$ is the total number of locations. These are to be compared against the quantities $h_{l,m}(\bm{x})$, where $h_{l,m}$ are transfer functions that relate the state variables in the model to the quantities being measured, at each location. In our optimization design, the measured quantities are the values of $P_z$ for each neutrino energy, at two locations: (i) the center of the Sun, and (ii) the surface of Earth. At the Sun's center, the ``measurement" of $P_z$ can be compared directly against the model $P_z$ at the same location, rendering the transfer functions trivial: $h_0(\vec P) = P_z$ for each neutrino.  On the other hand, since our model grid does not extend beyond the solar surface to the Earth, the $P_z$ measurement at Earth is compared against an extrapolated $P_z$ value derived from the Polarization vector at the Sun's surface, as shown in Eq.~\eqref{eq:Pzearth}.  In other words, measuring $P_z$ at Earth is equivalent to measuring a linear combination of $P_z$ and $P_x$ at the surface of the Sun. Therefore, the transfer function at the Sun's surface becomes:
\begin{equation} \label{eq:h}
h_{\odot s}(\vec P) = \left(1-\sin^22\theta\right) P_z- \cos2\theta \sin 2\theta \, P_x,
\end{equation}
for each neutrino energy\footnote{Note that in our previous papers~\cite{armstrong2017optimization,armstrong2022inferenceA,armstrong2022inferenceB,armstrong2020inference,rrapaj2021inference}, the transfer function was always trivial, since we simply employed simulated data at the far boundary of the model domain.}. The measurement term can then be written as
\begin{widetext}
\begin{equation} \label{eq:actionmeas}
\begin{split}
  A_{\text{meas}} =& \frac{1}{N_{\text{meas}}} \sum_{k=1}^{N_\nu} \left[\left(P_{z,k}^\text{meas} \left(0\right) - P_{z,k}\left(0\right)\right)^2 + \left(P_{z,k}^\text{meas}\left(R_\odot + L\right) - h_{\odot s}\left( \vec{P}_k\left(R_\odot\right) \right)\right)^2\right]
\end{split}
\end{equation}
\end{widetext}
where $k=\{1,\ldots,N_\nu\}$ are the neutrino energy bins, $\vec{P}_k$ is the polarization vector for the $k^\text{th}$ energy bin in the model equations of motion (Eq.~\eqref{eq:model1}), with components $P_{x,k}, P_{y,k}, P_{z,k}$, and $P_{z,k}^\text{meas}$ is the measurement of $P_z$ at the specified location, directly associated with the survival probability at that location (Eq.~\eqref{eq:PzPs}). $R_\odot$ and $L$ are the solar radius and the earth-sun distance, respectively.



\subsection{\textbf{Identifying a lowest minimum of the action}} \label{sec:methodAnnealing}

The action surface of a nonlinear model will be non-convex.  To identify a lowest minimum, we perform an iterative annealing in terms of the ratio of model and measurement error, $R_f$ and $R_m$, respectively\footnote{More generally, $R_m$ and $R_f$ are inverse covariance matrices for the measurement and model errors, respectively.  In this paper the measurements are taken to be mutually independent, rendering these matrices diagonal.}~\cite{ye2015systematic}.  It works as follows.

We define $R_m$ to be a constant (in this paper it is 1.0), and $R_f$ as: $R_f = R_{f,0}\alpha^{\beta}$, where $R_{f,0} = 10^{-3}$, $\alpha = 2.0$, and $\beta$ -- the annealing parameter -- is initialized at zero.  Relatively free from model constraints, the action surface is convex.  Then we increase $\beta$ recursively, each time recalculating the action, toward the deterministic limit of $R_f \gg R_m$.  The aim is to remain sufficiently near to the lowest minimum so as not to become trapped in a local minimum as the model dynamics resolve the surface\footnote{The complete procedure -- a variational approach to extremization and an annealing method to identify a lowest minimum of the cost -- is termed variational annealing (VA).}.

\section{The task for optimization} \label{sec:expers}

Our model beams were assigned energies corresponding to the experimentally detected neutrino energies.  We placed two constraints on each beam.  One was an assumed "measurement" of \(P_z\) at the center of the Sun: 1.0 for each beam, or: pure $\nu_e$ flavor.  The other measurement was \(P_{\text{survival}}\), the electron flavor survival probability of neutrinos measured at Earth's surface, related to the $P_z$ at earth through Eq.~\eqref{eq:PzPs}. 

\begin{figure}[htb] 
  \includegraphics[width=0.47\textwidth]{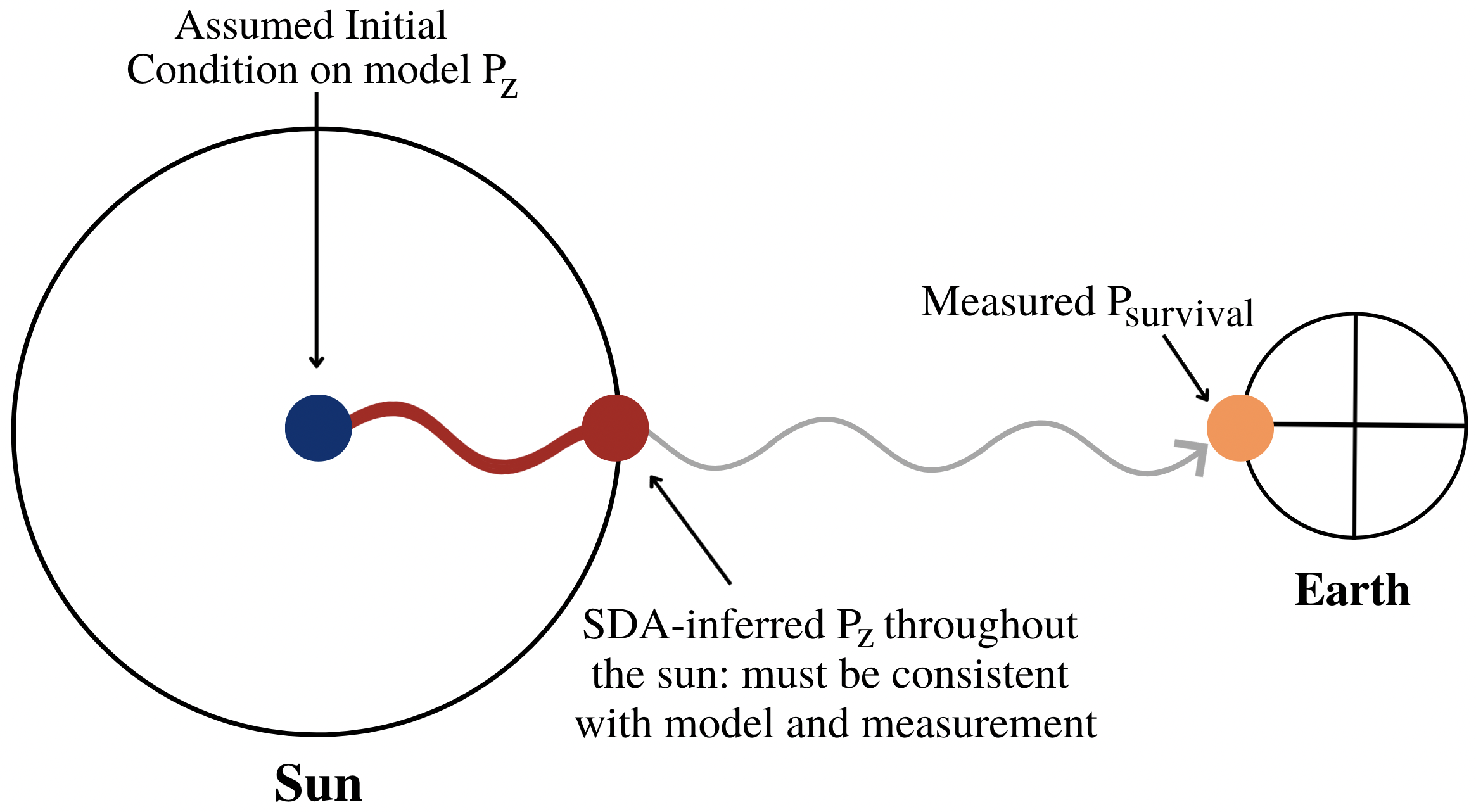}
  \caption{\textbf{Schematic of the inference task.} The constraints provided to the procedure are an assumed initial condition at the solar center (blue circle) and a measured survival probability at Earth (yellow circle).  Meanwhile, the model consists of the potential \(V(r)\) and flavor transformation dynamics (Eq.~(\ref{eq:model1})) through the Sun.  The inference task is to consider both the constraints and the model, to predict the flavor evolution within the Sun (red region), while also accounting for the transformation (grey wave) of the Earth-based measurement (yellow circle) to the solar surface (red circle).}
 \label{fig:schematic}
\end{figure}

The task for the procedure, depicted in Fig.~\ref{fig:schematic}, was to take those two constraints, together with the potential \(V(r)\) from the standard solar model~\cite{bahcall2005new} and the dynamics of flavor evolution (Eq.~(\ref{eq:model1})), to find a solution -- in terms of the radial evolution of the polarization vectors -- that is consistent with model and data.  We expected this inference task to be a challenge, specifically for estimating the polarization vector at the solar surface. 
\noindent From Eq.~\eqref{eq:connection}, one can see that
a single measured value of survival probability at Earth corresponds to a (\(P_x\),\(P_z\)) pair at the solar surface. In principle, many possible pairs could satisfy that relation. The SDA procedure is tasked with finding a pair that also obeys model dynamics (Eq.~(\ref{eq:model1})). 

\subsection{Details} \label{sec:SDAdetails}

The procedure is given full knowledge of model parameter values (except for the parameter estimation of Section~\ref{sec:paramEst}), all listed in Table~\ref{table:params}.  The three energies $E_B,i$ correspond to three discrete bins in the Borexino experiment~\cite{BOREXINO:2018ohr}; SNO employed an energy range, which we discretized into eight values~\cite{SNO:2009uok}.  To assign a unique value of the matter potential \(V(r)\) at each radial location, we used a piecewise linear interpolation: the \(V(r)\) term adopted from the solar model in Ref.~\cite{bahcall2005new} contains values at 1,219 discrete radial locations, while our model contains 121,901\footnote{Our choice to increase the number of discrete locations by a factor of 100 was motivated by our aim to resolve the oscillations; namely, the step size had to be sufficiently small.}.




\setlength{\tabcolsep}{5pt}
\begin{table*}[htb]
\caption{\textbf{Model parameters taken to be known.} The $E_{B,i}$ and $E_{SNO,i}$ correspond to the Borexino and SNO beams, respectively.  The $P_{\text{survival},B,i}$ are the measured survival probabilities for each Borexino energy bin \cite{BOREXINO:2018ohr}.  The $c_i$ values and uncertainties are used in Eq. \eqref{eq:SNO} to compute the survival probabilities for the SNO beams.}
\centering
\begin{tabular}{| c | c | c | c |} \toprule
\hline
 \textit{Parameter} & \textit{Value [MeV]} & \textit{Parameter} & \textit{Value} \\\midrule \hline
 $E_{B,1}$ & 7.4 & $P_{\text{survival},B,1}$ & $0.39\pm0.09$\\
 $E_{B,2}$ & 8.1 & $P_{\text{survival},B,2}$ & $0.37\pm0.08$\\
 $E_{B,3}$ & 9.7 & $P_{\text{survival},B,3}$ & $0.35\pm0.09$ \\ \midrule \hline
 $E_{SNO,1}$ & 8.5 & $c_0$ & $0.3435^{+0.0233}_{-0.0208}$\\
 $E_{SNO,2}$ & 9.0 & $c_1$ & $0.00795^{+0.00838}_{-0.00817}$ $\text{MeV}^{-1}$\\
 $E_{SNO,3}$ & 9.5 & $c_2$ & $-0.00206^{+0.00336}_{-0.00336}$ $\text{MeV}^{-2}$\\
 $E_{SNO,4}$ & 10.0 & &\\
 $E_{SNO,5}$ & 10.5 & &\\
 $E_{SNO,6}$ & 11.0 & &\\
 $E_{SNO,7}$ & 11.5 & &\\
 $E_{SNO,8}$ & 12.0 & &\\
 \bottomrule \hline
\end{tabular} 
\label{table:params}
\end{table*}




The optimization was performed by the open-source Interior-point Optimizer (Ipopt)~\cite{wachter2009short}.  Ipopt employs a Hermite-Simpson method of discretization and a constant step size.  We employed 121,901 steps and a step size of $\delta r$ of 2.85237.  The discretization of state space, calculations of the model Jacobean and Hessian matrices, and the annealing procedure are performed via a Python interface~\cite{minAone} that generates C code to be read by Ipopt.  Simulations were run on a computing cluster equipped with 201 GB of RAM and 24 GenuineIntel CPUs (64 bits), each with 12 cores.  

To compare solutions to the model dynamics, we generated a simulation via forward integration, with all beams initialized at [$P_x$,$P_y$,$P_z$]=[0,0,1] at the solar core, and using the same discretized grid as the optimization.  This integration was performed by Python's odeINT package, which uses an adaptive step.  Our complete procedure, including forward-integration codes, codes to interface with Ipopt, instructions for designing and running experiments on our supercomputing cluster, and examples for new users, can be found in a publicly available repository~\cite{github}.

For all experiments to be described in Section~\ref{sec:results}, four independent paths were initialized randomly.  That is, each initialization consisted of as many random choices as there are dimensions in the model\footnote{As noted, generally the dimensionality of the action surface is $((D + p) \times (N+1))$, where $D$, $N+1$, and $p$ are the number of state variables, discretized model locations, and parameters, respectively.  All optimization procedures described in this paper, however, contained at most one unknown parameter, which was assumed to have the same value at all (N+1) discretized locations.  Thus, the dimensionality of these procedures was: $D \times (N+1) + p$.}.  The user-defined search range for state variables ($P_x$,$P_y$, and $P_z$) was: [-1.0:1.0], the full dynamical range for each.  For the search range used in parameter estimation, see Section~\ref{sec:paramEst}.

\section{Results} \label{sec:results}

Key findings are as follows:
\begin{itemize}
    \item When the MSW transition within the Sun is included in the model dynamics, the procedure finds a solution consistent with both model dynamics and measured survival probabilities $P_\text{survival}$ at Earth, within the published experimental errors on $P_\text{survival}$, for both SNO and Borexino data.  (Alternatively, when the MSW transition is ignored -- or, \(V(r)\) is set to zero -- a solution compatible with both model and measurements is not found.)
    \item A preliminary parameter estimation shows that the measured survival probabilities contain information about model mixing angle $\theta$ (Eq.~(\ref{eq:model1})).
\end{itemize}

\begin{figure*}[htb] 
  \includegraphics[width=0.9\textwidth]{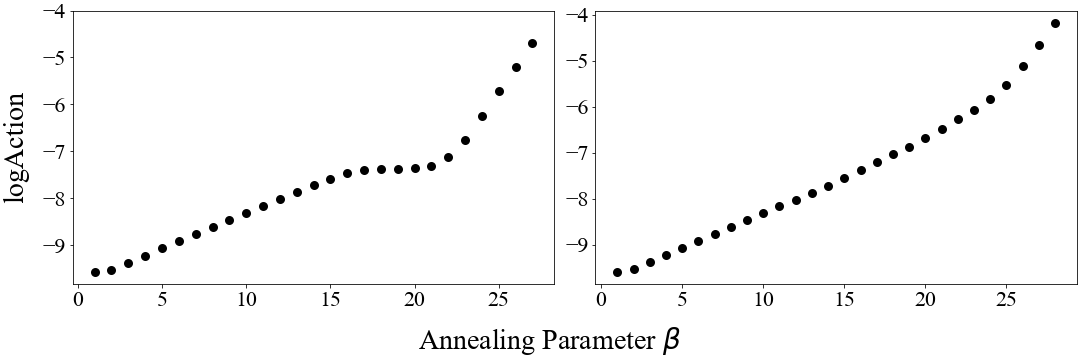}
  \caption{\textbf{Logarithmic plot of the action \(A_0\) as a function of annealing parameter \(\beta\).}  \textit{Left}: \(V(r)\) is taken from the standard solar model~\cite{bahcall2005new}.  At $\beta=0$ the action begins to increase as model weight increases, then plateaus -- at $\beta \sim 16\text{--}20$ -- as a solution is found that is consistent with both model and measurement (the increase in action beyond $\beta \sim 20$ is due to discretization error at high model weight).  \textit{Right}: \(V(r)\) is set to zero.  Now the action increases exponentially, indicating a failure to reconcile model with measurement.  (See Ref.~\cite{armstrong2020inference} for a detailed study of the action($\beta$) plot.)}
\label{fig:VBaction}
\end{figure*}

\begin{figure*}[htb]
\begin{center}
 \includegraphics[width = 0.49\textwidth]{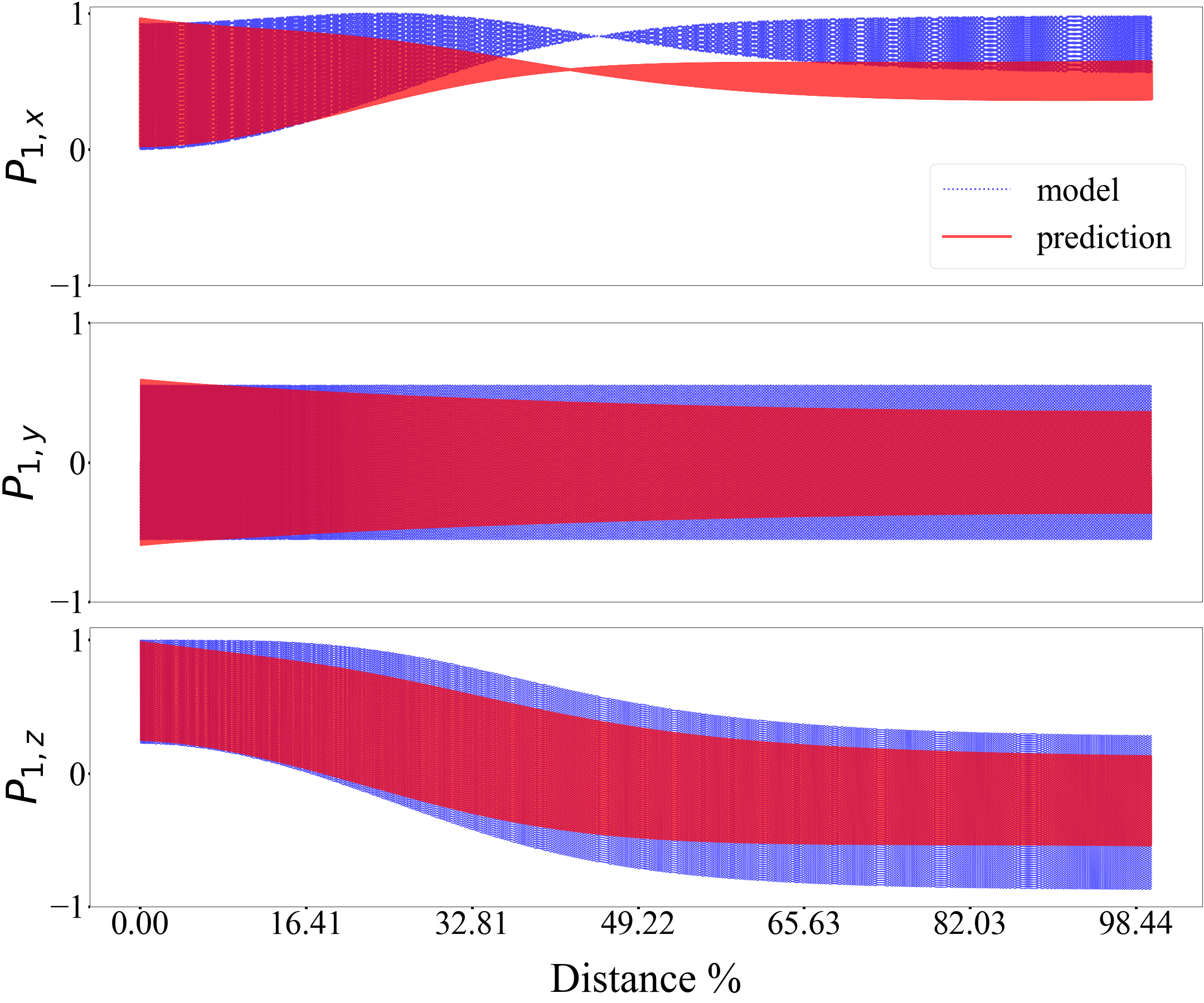}
 \includegraphics[width = 0.49\textwidth]{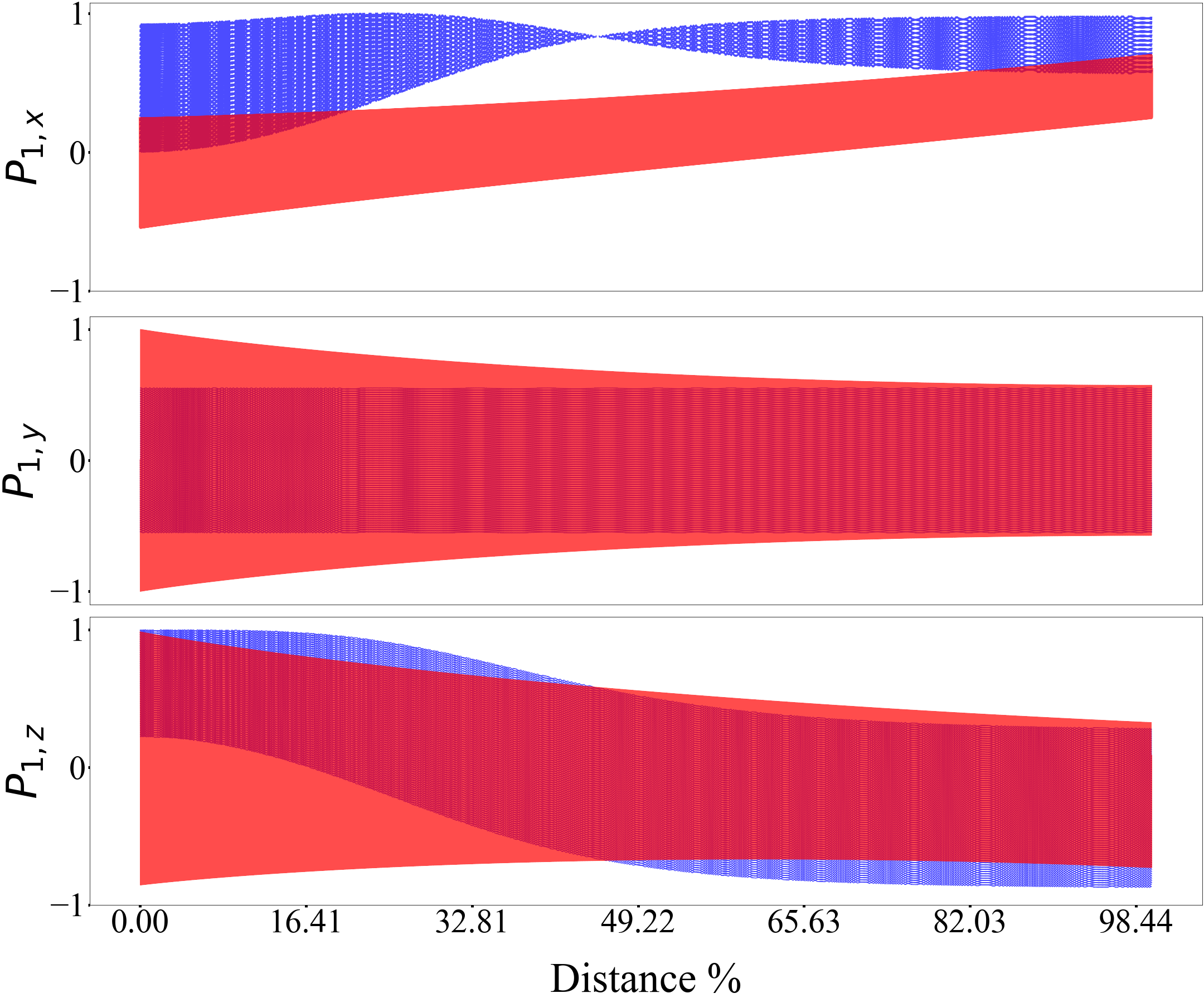}
 \caption{\textbf{True (blue) and predicted (red) state variable evolution, given model dynamics (Eq.~(\ref{eq:model1}) and (\ref{eq:h})) and Earth-based survival probability.}  \textit{Left}: for matter potential \(V(r)\) taken from the standard solar model; \textit{right}: with \(V(r)\) set to zero.  From top: $P_{1,x}$, $P_{1,y}$ and $P_{1,z}$, for the first of three energy beams corresponding to the Borexino data.  The units of distance are in percentage of solar radius $R_\odot$.  Both left and right panels correspond to a value of annealing parameter $\beta$ of 20.  For the result at left, $\beta=20$ lies on the "plateau" of Fig.~\ref{fig:VBaction}, left panel -- a solution consistent with both model and measurement.  At right, a solution compatible with model dynamics is not found.  These results are representative of all beams across both Borexino and SNO data sets, and for cases in which we added to the measurements the published maximum values of experimental error (not shown).}    
 \label{fig:whole_trajectory_Pxyz}
\end{center}
\end{figure*}

\subsection{\textbf{State prediction without parameter estimation}}

The left panel of Fig.~\ref{fig:VBaction} shows the logarithm of the action as a function of annealing parameter $\beta$, over the course of annealing, for the case in which the matter potential \(V(r)\) is taken from the standard solar model (Section~\ref{sec:method}), a scenario wherein the electron number density effects a significant MSW transition.  

Beginning at $\beta = 0$, the action increases as the weight of model error is increased.  Gradually, however, it levels off.  This "plateau," around  $\beta \sim 16$ to 20, indicates that a solution consistent with both model and measurements has been found\footnote{The further increase in the action beyond $\beta \sim 21$ is most likely due to discretization error, as the optimization technique uses a different method of discretization compared to the forward integration.} (for a detailed study of these action($\beta$) plots, see Ref.~\cite{armstrong2020inference}).  All four randomly-initialized paths converged to an identical solution.

Indeed, the predictions of state variable evolution well match the true model evolution.  The left panel of Fig.~\ref{fig:whole_trajectory_Pxyz} shows the true (blue) versus predicted (red) state variable evolution of $P_x$ (top), $P_y$ (middle), and $P_z$ (bottom), for the first of the three beams corresponding to the Borexino data; the result is representative of all beams in both SNO and Borexino models.  

\begin{figure*}[htb]
 \includegraphics[width = 19 cm, height = 8 cm]{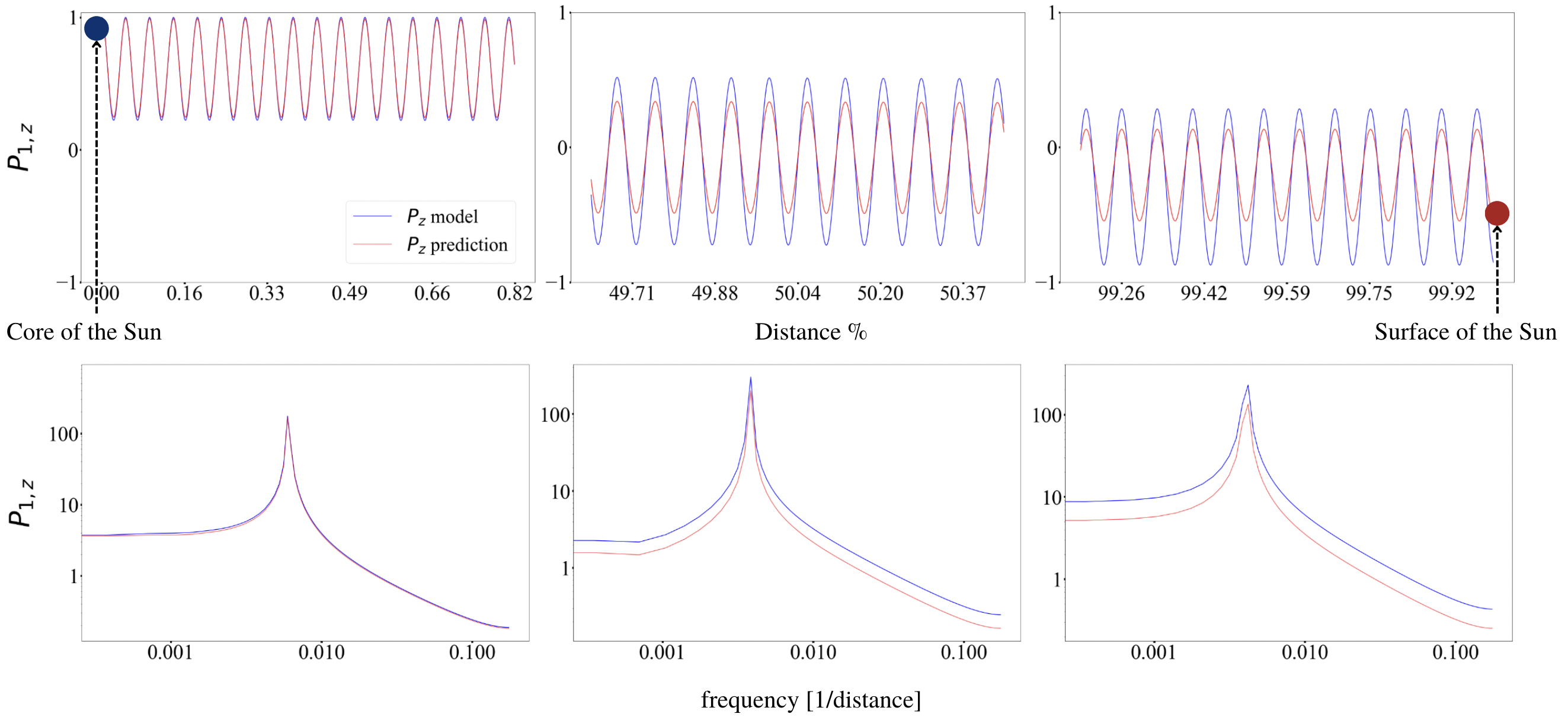}
 \caption{\textbf{Details of the predictions shown in Fig.~\ref{fig:whole_trajectory_Pxyz}.}  \textit{Top}: from left, three 1,000-point segments of the $P_z$ evolution, beginning at the solar center (far left) -- where the blue dot denotes the assumed initial condition, and ending at the solar surface (far right) -- where the red dot denotes the location where $P_x$ and $P_z$ must be consistent with the measurement at Earth (Earth is not depicted).  \textit{Bottom}: Fast Fourier decomposition for the respective regions at top, showing that the oscillation frequency is well predicted throughout the Sun.}
 \label{FT_first_middle_last1000}
\end{figure*}

Fig.~\ref{FT_first_middle_last1000} shows detail that cannot be discerned by eye on Fig.~\ref{fig:whole_trajectory_Pxyz}.  At top are three segments of the evolution of $P_z$: the first thousand steps, beginning at the solar center (left), middle thousand (middle), and final thousand ending at the solar surface (right); each section is roughly one hundredth of the full 121,901-step series.  The blue solid dot at far left denotes the location of the assumed initial conditions on $P_z$, and the red dot at far right indicates the first location of prediction, given the survival probability measured at Earth (Earth is not depicted in the figure).  

At bottom are the Fourier transforms of $P_z$ corresponding to each segment. The predicted oscillation frequency throughout the Sun is perfect, to within the resolution permitted by the density of sampled locations. These results are insensitive to the addition of the published maximum experimental errors on survival probability, for both SNO and Borexino.

As a test that the procedure recognizes consistency between measurement and model, we set the matter potential \(V(r)\) to zero and repeated the optimization.  Setting \(V(r)\) to zero essentially tells the procedure that the electron number density inside the Sun is not sufficiently high to effect an appreciable MSW transition.  

The right panel of Fig.~\ref{fig:VBaction} shows the resulting plot of action-versus-$\beta$.  In contrast to our original result, here the action never attains the signature "plateau" indicative of a successful optimization.  Rather, it increases exponentially.  The corresponding predictions appear in the right panel of Fig.~\ref{fig:whole_trajectory_Pxyz}: agreement with true (blue) model is poor.  As with the original experiments, all four randomly-initialized paths converged to this solution.  Together, Figures ~\ref{fig:VBaction} and \ref{fig:whole_trajectory_Pxyz} and convince us that the MSW effect within the Sun cannot be neglected if one seeks to account for the measured Earth-based survival probabilities. 

We conducted a further test that the SDA procedure recognizes consistency between measurement and model.  As noted, one vital component of a successful prediction was a ($P_x,P_z$) pair at the solar surface that was consistent with both the survival probability measured at earth \textit{and} the model dynamics (Eq.~(\ref{eq:model1})), together with the Earth-Sun transformation function of Eq.~(\ref{eq:h}).  With this in mind, we replaced the true published measured survival probabilities with "test" survival probabilities that are not physically possible; specifically: values above 1.0 and below 0.0.  For these cases, the action($\beta$) plot behaved similarly to the right panel of Fig.~\ref{fig:VBaction}: an exponential increase (not shown).  That is, the model was unable to find a solution that satisfied both model and measurements -- as is the expected outcome.

\subsection{\textbf{State prediction with simultaneous parameter estimation}} \label{sec:paramEst}

Adding parameter estimation to the inference task renders it significantly more challenging~\cite{ruiz2013estimating,carrassi2011state}.  The root of the difficulty is that parameters -- unlike state variables -- do not obey a known dynamical law, so there is no straightforward way to correlate state variable evolution with parameter estimate error\footnote{To first order, that correlation will be proportional to the model's Jacobian with respect to the parameter in question.}.  With that in mind, we took the preliminary step of setting the mixing angle $\theta$ of Eq.~(\ref{eq:model1}) as an unknown parameter to be estimated.  In these tests, the true value of $\theta$ was 0.58, and the permitted search range was 0.001 to 1.571.

\begin{figure}
    \centering
    \includegraphics[width=9cm]{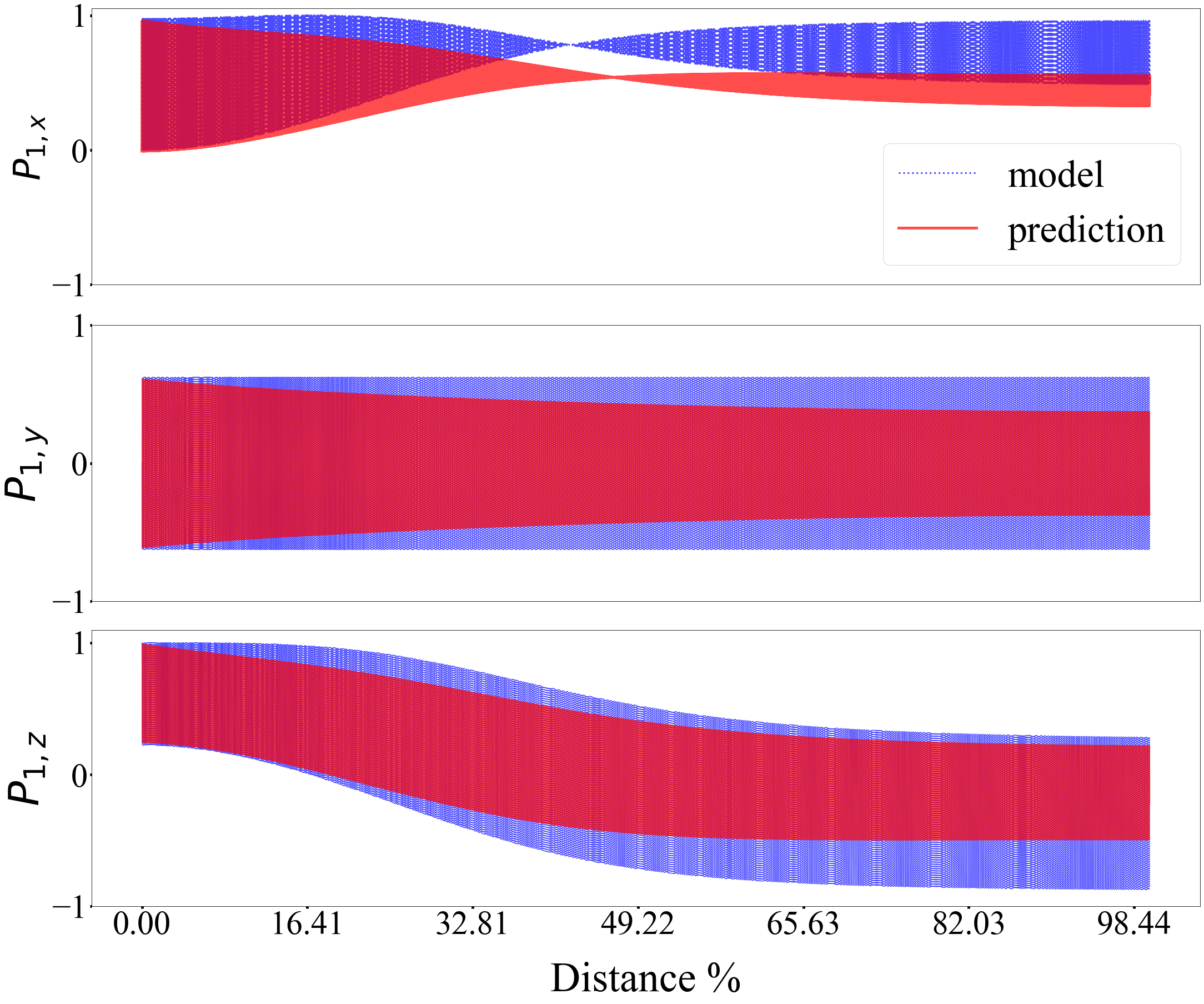}
    \caption{\textbf{State predictions akin to Fig.~\ref{fig:whole_trajectory_Pxyz}, now with mixing angle $\theta$ recast as an unknown parameter to be estimated -- where the Earth-Sun transformation has been omitted in the interest of easing computational expense.}  These predictions correspond to an estimate of $\theta$ of 0.59 radians (true: 0.58), where the permitted search range was [0.001,1.571].  The solution corresponds to a value of $\beta$ of 20.}
    \label{fig:paramplot1}
\end{figure}

Our first parameter attempt failed (not shown), and to ascertain whether computational expense might be the cause, we temporarily removed the Earth-to-Sun transformation of Eq.~(\ref{eq:h}) from the model dynamics.  To be clear: our aim here was to offset the increased computational expense incurred with parameter estimation by decreasing the complexity of the equations of motion, to identify whether computational expense was the culprit in the failed first attempt.  Across four randomly-initialized paths, the resulting estimates of $\theta$ were, respectively: [1.57, 0.98, 0.01, 0.59] radians.  That is, one of the four paths identified a solution near the true value of 0.58.  Fig.~\ref{fig:paramplot1} shows the corresponding state prediction for that estimate.

\begin{figure}
    \centering
    \includegraphics[width=9cm]{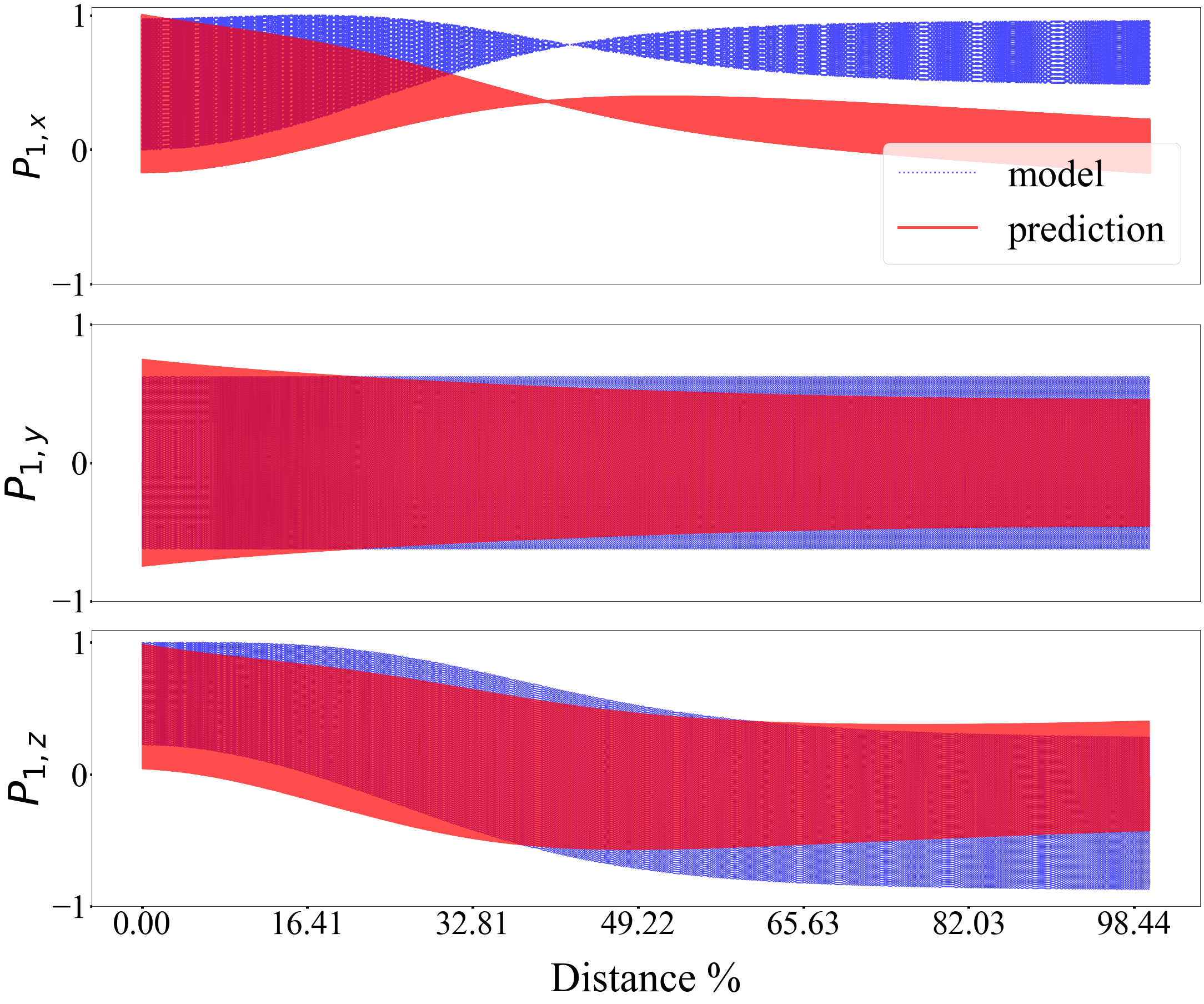}
    \caption{\textbf{State predictions akin to Fig.~\ref{fig:whole_trajectory_Pxyz}, again with mixing angle $\theta$ recast as an unknown parameter to be estimated, and including the Earth-Sun transformation.}  Importantly, in this version of the experiment, the measurements were taken to correspond not only to the ($P_x$,$P_z$) pair at the solar surface $R_\odot$, but rather to the average of those two values over the final 1,000 discretized model locations in the Sun.  See text for explanation.  One out of four paths converged to this solution, which corresponds to a value of  $\theta=0.56$, taken at a value of $\beta$ of 20.}
    \label{fig:paramplot2}
\end{figure}

Encouraged by this mild success, we restored the Earth-Sun transformation (Eq.~(\ref{eq:h})) -- and in addition, we amended the requirement regarding a match to the Earth-based measurement of survival probability.  Specifically, instead of using only the final values of \(P_x\) and \(P_z\) at the solar surface, we averaged \(P_x\) and \(P_z\) over the last 1000 radial locations in the model and compared those averages to the measured survival probability, using the Sun-Earth transformation described earlier.  The motivation for this change was that, without a fixed mixing angle, the rapid oscillations in the predicted path could, in principle, be out of phase with the model path.  With this change, the results improved: the estimates of $\theta$ across four independent paths were: [0.79, 0.78, 0.56, 0.89] radians, respectively.  Fig.~\ref{fig:paramplot2} shows the corresponding state predictions for the closest estimate of 0.56.


%


There is much honing to be done to yield reliable results from parameter estimation.  We need to ascertain why the problem invites multiple solutions across independent paths searched, and aim to increase the fraction of successful paths.  We note, however, that based on this preliminary study, inference can indeed extract information about the mixing angle in Eq.~(\ref{eq:model1}) from Earth-based measurements. 

\section{Conclusion}

For the first time we have applied inference to astrophysical neutrino data.  The ability of the SDA procedure to identify state predictions consistent with both model and Earth-based measurements is encouraging, particularly given the multiple unphysical solutions at the solar surface that those measurements permit.  Further, while the parameter estimation needs honing, the results presented in Section~\ref{sec:paramEst} indicate that inference-based procedures do possess the capability to recognize the dependence of Earth-based flavor survival probability measurements on the neutrino mixing angle $\theta$.
Finally, for the interested reader, the Appendix addresses a topic noted briefly earlier: the phenomenon of decoherence through neutrino propagation.

\begin{acknowledgments}
C.~L., E.~A., A.~B.~B., A.~A., M.~S., and I.~S. acknowledge NSF grant PHY-2139004.  S.~W. acknowledges the NSF summer Research Experience for Undergraduates program. The work of A.~V.~P. was supported by the U.S. Department of Energy under contract number DE-AC02-76SF00515. The work of A.B.B. was also supported by the NSF Grants No. PHY-2020275 and PHY-2108339.  As always, eternal thanks to the good people of Kansas and Doylestown, Ohio.
\end{acknowledgments}


\section*{Appendix: Coherent versus incoherent detection} \label{sec:decoherence}

In Section~\ref{sec:evol}, we wrote a transformation matrix to carry the polarization vector between the Earth and Sun.  { In doing so, we omitted an accounting of the loss of coherence in the neutrinos, due to the constituent mass eigenstates becoming separated in space on their way to the detector.  It might be valuable in the future to investigate whether including this decoherence in the model improves state and parameter estimation.  It must be noted that this loss of coherence is a well-understood phenomenon that has been described in the context of solar as well as supernova neutrinos (e.g.,~\cite{Dighe:1999id,deHolanda:2019tuf,Kersten:2015kio}). Here we provide a brief overview to the interested reader of how this decoherence may affect the final state that we measure at earth.}  

When a neutrino arrives at earth from a sufficiently distant source, 
 it is in general a coherent superposition of more than one mass (i.e., propagation) eigenstate. However, since the different propagation eigenstates have different masses and therefore propagate at different velocities, their wavepackets eventually become spatially separated---to the extent that any neutrino interaction in a detector would only involve the participation of \emph{one} of these mass eigenstates at a time.

Since the individual mass eigenstates themselves do not oscillate in time (only their coherent superpositions do), the probability of detecting a particular mass eigenstate in a certain flavor (e.g., $P_{1e} = \braket{\nu_e}{\nu_1}$) is independent of time or position, once the neutrino is propagating in vacuum (i.e., if there are no matter effects). For the purposes of detection, since only one mass eigenstate participates at a time (assuming sufficient wavepacket separation), one may treat the neutrino flux arriving at the earth as an incoherent mixture of mass eigenstates rather than as a coherent superposition. Mathematically, this can be described as follows.

Consider a single neutrino existing as a coherent superposition of mass eigenstates. Its density matrix in the mass basis may be written as follows:
\begin{equation}
    \rho_\nu = 
    \begin{pmatrix}
        n_{\nu_1} & \rho_{12} \\
        \rho_{12}^\star & n_{\nu_2}
    \end{pmatrix},
\end{equation}
where $n_{\nu_i} = \langle a^\dagger_{\nu_i}a_{\nu_i} \rangle$ are expectation values of the number operators for the mass eigenstates $i \in \{1,2\}$, and $\rho_{12}$ depends on the relative phase between the mass eigenstates. For a neutrino propagating in vacuum, the $n_{\nu_i}$ remain invariant in time, and only the off-diagonal entries are time-dependent. 

In contrast, for a neutrino that can be considered to have essentially devolved into an incoherent mixture of mass eigenstates, the density matrix is simply 
\begin{equation}
    \rho_\nu = 
    \begin{pmatrix}
        n_{\nu_1} & 0 \\
        0 & n_{\nu_2}
    \end{pmatrix}.
\end{equation}

In each case, one may ask the question of how much can be learned from a measurement (i.e., a detection). First, let us define the quantities $n_{\nu_e}$ and $n_{\nu_x}$, which are the expectation values of the flavor-basis number operators. These can be calculated as follows: 
\begin{equation}
\begin{split}
 n_{\nu_e} &= \langle a^\dagger_{\nu_e} a_{\nu_e} \rangle = \Tr{\rho_\nu \, a^\dagger_{\nu_e} a_{\nu_e}} \\
           &= n_{\nu_1} \cos^2{\theta} + n_{\nu_2} \sin^2{\theta} + 2\,\Re{\rho_{12}} \sin\theta \cos\theta, \\
 n_{\nu_x} &= \langle a^\dagger_{\nu_x} a_{\nu_x} \rangle = \Tr{\rho_\nu \, a^\dagger_{\nu_x} a_{\nu_x}} \\
           &= n_{\nu_1} \sin^2{\theta} + n_{\nu_2} \cos^2{\theta} - 2\,\Re{\rho_{12}} \sin\theta \cos\theta. \\
\end{split}
\end{equation}

Here, we have used the definitions from earlier in the section, and the familiar unitary transformation between the flavor and mass eigenbasis:
\begin{equation}
    \begin{pmatrix}
        a_{\nu_e} \\ a_{\nu_x}    
    \end{pmatrix}
    =
    \begin{pmatrix}
        \cos\theta & \sin\theta \\
        -\sin\theta & \cos\theta
    \end{pmatrix}
    \begin{pmatrix}
        a_{\nu_1} \\ a_{\nu_2}
    \end{pmatrix}.
\end{equation}

Suppose, for instance, one detects enough neutrinos at a certain energy so as to statistically obtain a sufficiently accurate determination of $n_{\nu_e}$ and $n_{\nu_x}$ at a given location. Then, in case of neutrinos that are still coherent superpositions by the time of arrival at the detector, one cannot uniquely determine $n_{\nu_1}$ and $n_{\nu_2}$ from $n_{\nu_e}$ and $n_{\nu_x}$, unless one also measures $\rho_{12}$, which is physically impossible. Instead, a possible workaround may be to measure $n_{\nu_e}$ and $n_{\nu_x}$ at \emph{multiple} locations, as was explored in Ref.~\cite{armstrong2022inferenceA} in the context of supernova neutrino detection.

On the other hand, if the neutrinos that arrive at the detector are an incoherent mixture, then $\rho_{12} = 0$, and a measurement of $n_{\nu_e}$ and $n_{\nu_x}$ at a single location is sufficient to uniquely determine $n_{\nu_1}$ and $n_{\nu_2}$.  

\end{document}